\documentclass[twocolumn,floatfix,superscriptaddress,showpacs,prl,noshowkeys]{revtex4}
\usepackage{graphicx}
\usepackage{color}
\DeclareGraphicsExtensions{.pdf}
\begin{document}
\title {Structural and electronic properties of germanene on MoS$_2$}

\author{L. \surname{Zhang}}
\thanks{These authors contributed equally to this work}
\affiliation{Physics of Interfaces and Nanomaterials, MESA+ Institute for Nanotechnology, University of Twente,  P.O. Box 217, 7500 AE Enschede, The Netherlands.}

\author{P. \surname{Bampoulis}}
\thanks{These authors contributed equally to this work}
\affiliation{Physics of Interfaces and Nanomaterials, MESA+ Institute for Nanotechnology, University of Twente,  P.O. Box 217, 7500 AE Enschede, The Netherlands.}

\author{A.N. \surname{Rudenko}}
\thanks{These authors contributed equally to this work}
\affiliation{Institute for Molecules and Materials, Radboud University,
Heijendaalseweg 135, 6525 AJ Nijmegen, The Netherlands}

\author{Q. \surname{Yao}}
\affiliation{Physics of Interfaces and Nanomaterials, MESA+ Institute for Nanotechnology, University of Twente,  P.O. Box 217, 7500 AE Enschede, The Netherlands.}

\author{A. \surname{van Houselt}}
\affiliation{Physics of Interfaces and Nanomaterials, MESA+ Institute for Nanotechnology, University of Twente,  P.O. Box 217, 7500 AE Enschede, The Netherlands.}

\author{B. \surname{Poelsema}}
\affiliation{Physics of Interfaces and Nanomaterials, MESA+ Institute for Nanotechnology, University of Twente,  P.O. Box 217, 7500 AE Enschede, The Netherlands.}

\author{M.I. \surname{Katsnelson}}
\affiliation{Institute for Molecules and Materials, Radboud University,
Heijendaalseweg 135, 6525 AJ Nijmegen, The Netherlands}

\author{H.J.W. \surname{Zandvliet}}
\email[Corresponding author: ]{h.j.w.zandvliet@utwente.nl}
\affiliation{Physics of Interfaces and Nanomaterials, MESA+ Institute for Nanotechnology, University of Twente,  P.O. Box 217, 7500 AE Enschede, The Netherlands.}

\date{\today}

\begin{abstract}
To date germanene has only been synthesized on metallic substrates. A metallic substrate is usually detrimental for the two-dimensional Dirac nature of germanene because the important electronic states near the Fermi level of germanene can hybridize with the electronic states of the metallic substrate. Here we report the successful synthesis of germanene on molybdenum disulfide (MoS$_2$), a band gap material. Pre-existing defects in the MoS$_2$ surface act as preferential nucleation sites for the germanene islands. The lattice constant of the germanene layer (3.8 $\pm$ 0.2 \AA) is about 20\% larger than the lattice constant of the MoS$_2$ substrate (3.16 \AA). Scanning tunneling spectroscopy measurements and density functional theory calculations reveal that there are, besides the linearly dispersing bands at the $K$ points, two parabolic bands that cross the Fermi level at the $\Gamma$ point. 
\end{abstract}

\pacs{73.22.-f, 81.05.Zx, 68.37.Ef}

\maketitle

The discovery that graphene, a single layer of $sp^{2}$ hybridized carbon atoms arranged in a honeycomb registry, is stable has resulted in numerous intriguing and exciting scientific breakthroughs \cite{een,twee}. The electrons in graphene behave as relativistic massless fermions that are described by the Dirac equation, i.e. the relativistic variant of the Schr\"{o}dinger equation. One might anticipate that elements with a similar electronic configuration, such as silicon (Si), germanium (Ge) and tin (Sn), also have a ``graphene-like'' allotrope. Unfortunately, silicene (the silicon analogue of graphene), germanene (the germanium analogue of graphene) and stanene (the tin analogue of graphene) have not been found in nature and therefore these two-dimensional (2D) materials have to be synthesized. Theoretical calculations have revealed that the honeycomb lattices of the ``graphene-like'' allotropes of silicon and germanium are not fully planar, but slightly buckled \cite{drie,vier}. The honeycomb lattices of these 2D materials consist of two triangular sub-lattices that are slightly displaced with respect to each other in a direction normal to the honeycomb lattice. Despite this buckling the 2D Dirac nature of the electrons is predicted to be preserved \cite{drie,vier}. Another salient difference with graphene is that silicene and germanene have a substantially larger spin-orbit gap than graphene ($<$0.05 meV). Silicene's spin-orbit gap is predicted to be 1.55 meV, whereas the predicted spin-orbit gap of germanene is even 23.9 meV. This is very interesting because graphene and also silicene and germanene are in principle 2D topological insulators and thus ideal candidates to exhibit the quantum spin Hall effect. The interior of a 2D topological insulator exhibits a spin-orbit gap, whereas  topologically protected helical edge modes exist at the edges of the material \cite{vijf,zes}. The two topologically protected spin-polarized edge modes have opposite propagation directions and therefore the charge conductance vanishes, whereas the spin conductance has a non-zero value.

In the past few years various groups have successfully synthesized silicene \cite{zeven,acht,negen} and germanene \cite{tien,elf,twaalf,dertien} on a variety of substrates. To date germanene has only been grown on metallic substrates, such as Pt(111) \cite{tien}, Au(111) \cite{elf}, Ge$_2$Pt \cite{twaalf,veertien} and Al(111) \cite{dertien}, which might hinder a proper decoupling of the key electronic states of germanene near the Fermi level from the underlying substrate. Only Bampoulis \emph{et al.} \cite{twaalf} managed to resolve the primitive cell of the buckled honeycomb structure of germanene. 

Here we report the growth of germanene on a band gap material, namely MoS$_2$. We found that the germanene layer, which is only weakly coupled to the MoS$_2$ substrate, exhibits a clear V-shaped density of states. These experimental observations are in very good agreement with density functional theory calculations. The synthesis of germanene on MoS$_2$ is a first step towards future germanene-based device applications. 

The scanning tunneling microscopy (STM) and spectroscopy (STS) experiments have been performed at room temperature with an ultra-high vacuum STM (Omicron STM-1). The base pressure of the ultra-high vacuum system is 3 $\times$ 10$^{-11}$ mbar. We have used electrochemically etched tungsten STM tips. The MoS$_2$ samples were freshly cleaved from synthesized 2H-MoS$_2$ (acquired from 2D Semiconductors) and mounted on a Mo sample holder and subsequently introduced into the ultra-high vacuum system. Ge was deposited onto the MoS$_2$ substrate, which was held at room temperature, by resistively heating a clean Ge(001) wafer at $\sim$1150 K. The Ge wafer was located at a distance of $\sim$10 mm from the MoS$_2$ substrate. Prior to the deposition the Ge(001) wafer was cleaned by a pre-anneal at 700 K for about 24 h followed by several cycles of argon ion bombardment at 800 eV at room temperature and annealing at 1100 K \cite{vijftien}. After the deposition of germanium, the MoS$_2$ sample was immediately inserted into the STM. The deposition rate was estimated by analyzing several STM images after the deposition of sub-monolayer amounts of germanium on the MoS$_2$ substrate.

The density functional theory (DFT) calculations were performed using the projected augmented wave (PAW) formalism \cite{zestien} as implemented in the Vienna Ab Initio Simulation Package (VASP) \cite{zeventien,achttien}. Exchange-correlation effects were taken into account within the dispersion-corrected nonlocal optB88-vdW functional \cite{negentien}. An energy cutoff of 600 eV for the plane-waves and a convergence threshold of 10$^{-5}$ eV were employed. To avoid interactions between the supercells, a 25 \AA\quad thick vacuum slab was added in the direction normal to the surface. The Brillouin zone was sampled by a (4 $\times$ 4) k-point mesh. Structural relaxation was performed, while keeping the lowermost layer of sulfur atoms fixed, until the forces acting on the other atoms were less than 10$^{-2}$ eV/\AA. 

\begin{figure}
\includegraphics[width=8.5cm]{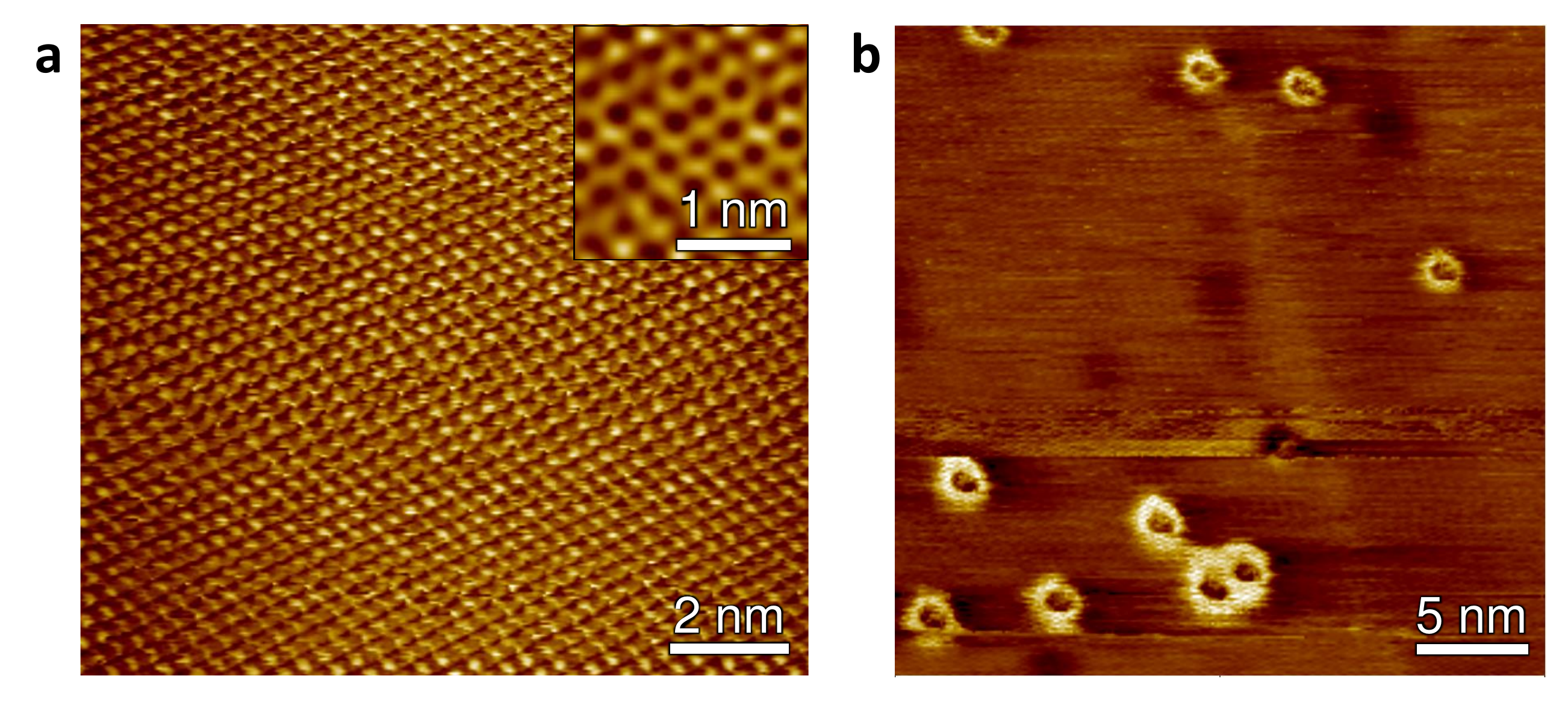}
\caption{
(Color on-line). {\bf (a)} Scanning tunneling microscopy image of the MoS$_2$ substrate. Inset: small scale image revealing the honeycomb structure of MoS$_2$ (obtained after a fast Fourier transform operation of the raw data). The set points are 0.2 V and 1 nA, respectively. 
{\bf (b)}
Scanning tunneling microscopy image of the MoS$_2$ surface after the deposition of $\sim$10 \% of a monolayer of germanium at room temperature. Small germanium islands are nucleated at pre-existing defect sites of the MoS$_2$ substrate. The set points are 0.5 V and 0.6 nA, respectively.
}
\label{fig1}
\end{figure}

\begin{figure}
\includegraphics[width=8.5cm]{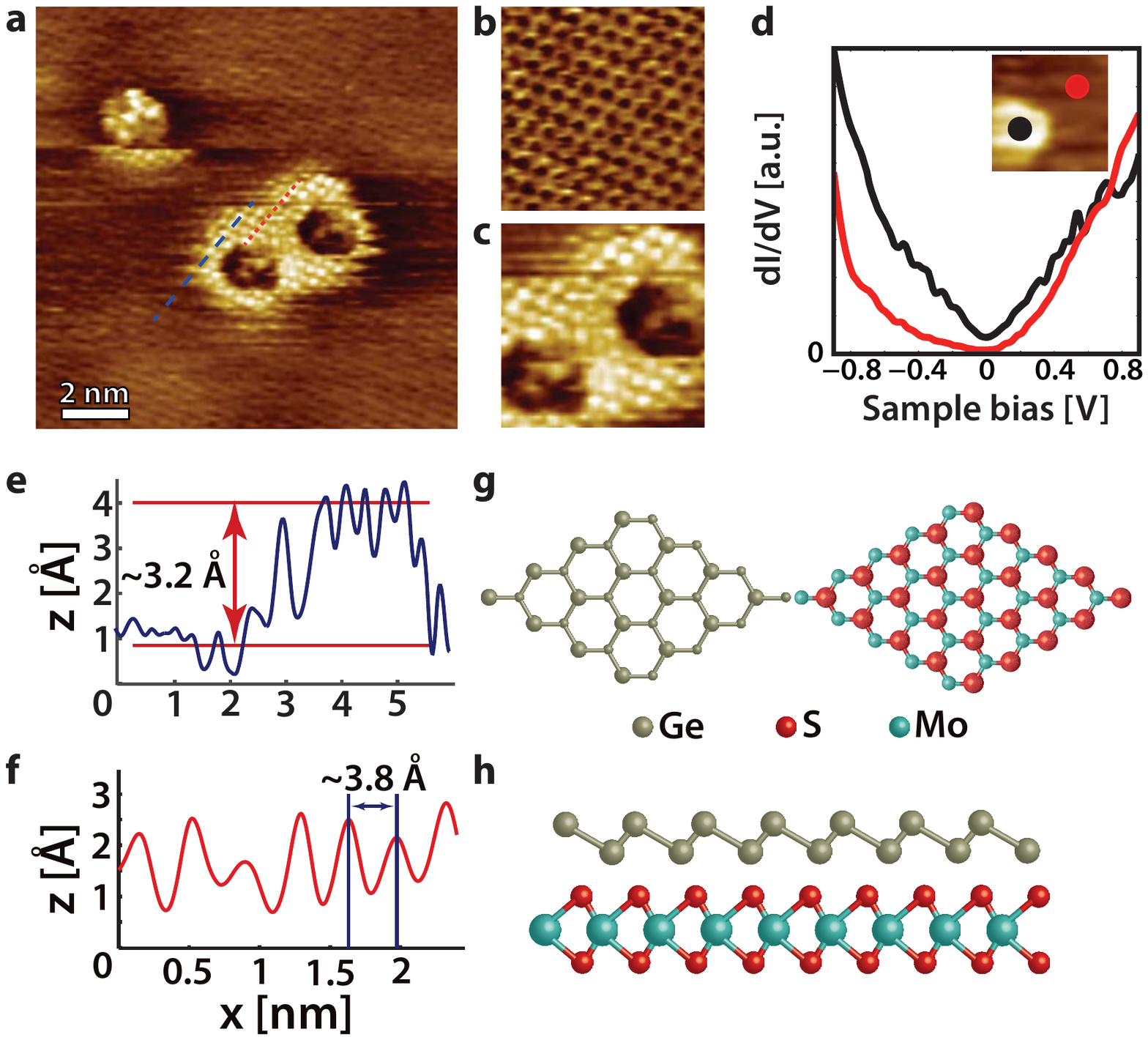}
\caption{
(Color on-line). {\bf (a)} Scanning tunneling microscopy image of the MoS$_2$/germanium substrate. The sample bias is 0.5 V and the tunneling current is 0.3 nA. 
{\bf (b)} A zoom-in on a bare MoS$_2$ area. The STM image reveals a honeycomb structure with a lattice constant of 3.15 $\pm$ 0.2 \AA, which corresponds to the lattice constant of MoS$_2$. The sample bias and the tunneling current are the same as in (a).
{\bf (c)} A zoom-in on the large germanene island of panel (a) reveals a hexagonal lattice with a lattice constant of 3.8 $\pm$ 0.2 \AA. The sample bias and the tunneling current are the same as in (a). 
{\bf (d)} Differential conductivity recorded on a germanene island (black curve) and the surrounding MoS$_2$ surface (red curve). The set points are 1 V and 0.3 nA. 
{\bf (e)} Line scan taken across the germanene island indicated with the blue dashed line in (a). The apparent height of the germanene islands is $\sim$3.2 \AA.
{\bf (f)} Line scan taken on top of the germanene island indicated with the red dashed line in (a). The measured lattice constant of the germanene island corresponds to 3.8 $\pm$ 0.2 \AA.
{\bf (g)-(h)} Ball and stick models of germanene and MoS$_2$. Top views (g) and side views (h). 
}
\label{fig2}
\end{figure}

Figure \ref{fig1}(a) shows an empty state STM image of the molybdenum disulfide (MoS$_2$) substrate. MoS$_2$ is a transition metal dichalcogenide. The elementary building block of a MoS$_2$ crystal is a tri-layer structure consisting of one close-packed Mo atomic layer encapsulated between two atomic layers of close-packed S atoms (see Figure \ref{fig2}(g)-(h)). The atoms within this layer are strongly bonded by covalent interactions, whereas the interactions between the MoS$_2$ layers are governed by weak Van der Waals forces. MoS$_2$ can easily be exfoliated and has a band gap that varies from 1.2 eV to 1.8 eV depending on its thickness. In most cases only the S-atoms of MoS$_2$ substrate are resolved in STM images, resulting in a triangular structure with a lattice constant of 3.16 \AA. In the inset of Figure \ref{fig1}(a) a high resolution STM image of the primitive honeycomb cell of MoS$_2$ is shown. The natural exfoliated MoS$_2$ surface usually contains some defects that are caused by a local variation in the stoichiometry, missing atoms or impurities. These defects can lead to $n$-type as well as $p$-type doping and are studied in depth \cite{twintig,twintig1,twintig2}. The defects affect the electronic structure of the MoS$_2$ in the vicinity of the defects and therefore they appear usually substantially larger in STM images than they really are (see Figure \ref{fig1}(b)). In Figure \ref{fig1}(b) a MoS$_2$ substrate is shown after the deposition of $\sim$10 \% of a monolayer of germanium at room temperature (here one monolayer corresponds to a (5 $\times$ 5) cell of germanene on a (6 $\times$ 6) cell of the MoS$_2$ substrate). Several germanium islands have been nucleated at pre-existing defects of the MoS$_2$ substrate. From image to image the shape and size of these germanium islands changes, which implies (1) attachment and detachment of germanium atoms at the edges of the germanium islands and (2) rapid diffusion of germanium atoms across the surface. Interestingly, there is a hexagonal shaped vacancy island in the center of all the nucleated germanium islands. 

\begin{figure}
\includegraphics[width=8.5cm]{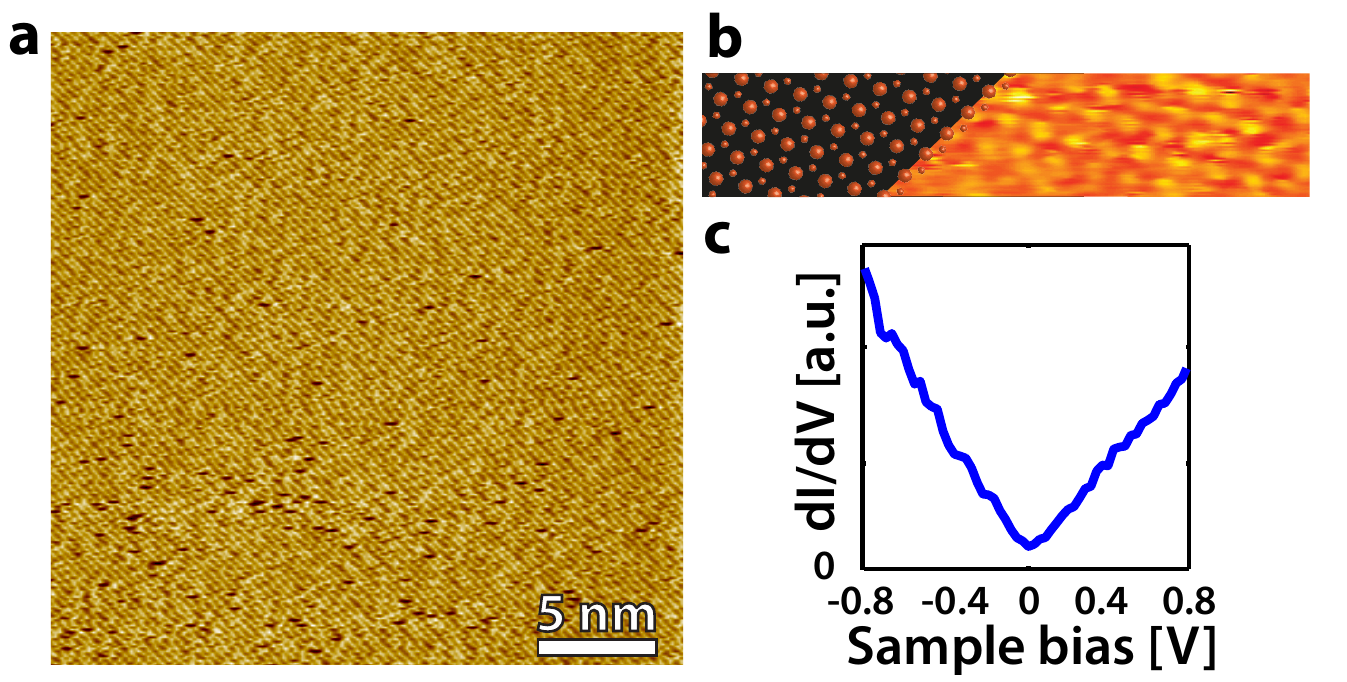}
\caption{
(Color on-line). {\bf (a)} A continuous film of germanene on MoS$_2$ after the deposition of about a monolayer of germanium. The sample bias and tunnel current are -1 V and 0.3 nA, respectively . 
{\bf (b)} Left: ball-and-stick model. Large (small) circles represent upward (downward) buckled germanium atoms. Right: High-resolution scanning tunneling microscopy image of germanene (only the upward buckled germanium atoms are resolved).
{\bf (c)} Differential conductivity recorded on the germanene film. The set points are -1.4 V and 0.6 nA, respectively.
}
\label{fig3}
\end{figure}

Figure \ref{fig2}(a) shows a small scale STM image of the MoS$_2$/germanium substrate. The frizzy appearance of the germanium islands is clearly visible. We cannot exclude that part of this dynamics is induced by the scanning process. In the upper left part of the image a germanium island changes abruptly in size. In Figures \ref{fig2}(b)-(c) zoom-ins of a germanium-free part of the MoS$_2$ substrate and a germanium island are shown, respectively. The lattice constant of the germanium islands is $\sim$ 20 \% larger than the lattice constant of the MoS$_2$ substrate, i.e. 3.8 $\pm$ 0.2 \AA \quad for germanium versus 3.16 \AA \quad for MoS$_2$. The germanium islands exhibit a threefold symmetry as the underlying MoS$_2$ substrate does. In addition, the high symmetry directions of the MoS$_2$ substrate and the germanium islands are aligned (see Figure \ref{fig2}(a)). 

Differential conductivity (\textit{dI/dV}) spectra recorded on a germanium island and the bare MoS$_2$ substrate are depicted in Figure 2(d). As expected the MoS$_2$ region has a band gap, whereas the differential conductivity of the germanium island reveals a well-defined V-shape around zero bias ($\left|V\right| <  $ 0.5 V). For small sample biases the differential conductivity is proportional to the density of states and therefore also the density of states of the germanium islands exhibits a V-shape, which is one of the hallmarks of a 2D Dirac system. Based on these observations we arrive at the conclusion that the germanium islands are actually germanene islands. This assignment does, however, not imply that these germanene islands also exhibit all the interesting and intriguing ``graphene" properties.

In figure \ref{fig2}(e)-(f) line scans taken across the germanene island (see Figure \ref{fig2}(a)) are depicted. The apparent height of the germanene island is $\sim$3.2 \AA. Since the electronic structure of germanene and MoS$_2$ are substantially different it is not appropriate to compare this experimentally determined step height with theoretical predictions. 

In Figure \ref{fig3}(a)-(b) an STM image is shown after the deposition of $\sim$1 monolayer of germanium. The germanene islands are now much larger and virtually the whole MoS$_2$ surface is covered with germanene. The hexagonal shaped vacancy islands have all disappeared. The periodicity of the structure in Figure \ref{fig3}(a)-(b) is again 3.8 $\pm$ 0.2 \AA. It is pointed out that the resolution in Figure \ref{fig3}(a) is somewhat lower than in the previous images. In Figure \ref{fig3}(b) only one of the two triangular sub-lattices is resolved. We did not find any evidence for the presence of a moir\'{e} pattern. Due to the 20 \% difference in lattice constants between germanene and MoS$_2$ one expects only very small moir\'{e} periodicities ($<$ 2 nm) \cite{twintig3}.

As shown in Figure \ref{fig3}(c) the differential conductivity has a well-defined V-shape and thus confirms our earlier assignment that we are dealing with germanene. The differential conductivity does, however, not completely vanish at the Dirac point indicating that the system is metallic and hence not an ideal 2D Dirac system. Free-standing low-buckled germanene is known to be a 2D Dirac system that shares many properties with its famous counterpart graphene \cite{twintig4}.  Here we are dealing with a germanene layer on a substrate and even though the interaction between germanene and MoS$_2$ is rather weak, the structural and electronic properties of germanene are affected by the substrate.

\begin{figure}
\includegraphics[width=8.5cm]{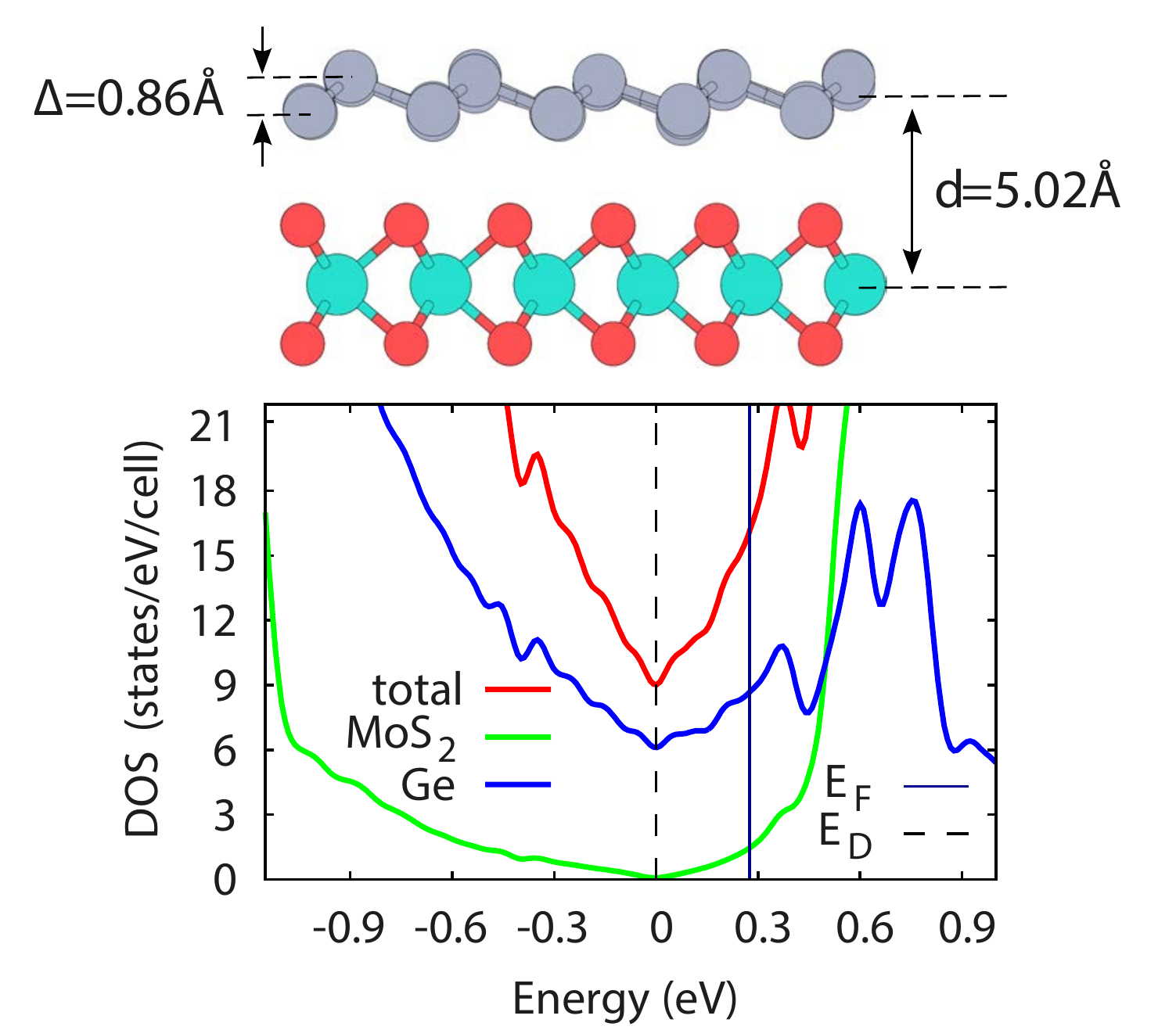}
\caption{
(Color on-line). Calculated total and partial densities of states of germanene on MoS$_2$. Vertical solid and dashed lines correspond to the Fermi energy and Dirac point, respectively.}
\label{fig4}
\end{figure}

To support our experimental observations of germanene, we have performed first-principles calculations using DFT. Unlike earlier DFT studies, where germanene has been considered on MoS$_2$ in its free-standing configuration with a lattice constant of ~3.97 \AA \cite{twintig5,twintig6}, we consider here germanene laterally contracted by $\sim$5 \% in accordance with the experimental observations. Particularly, we use a (5 $\times$ 5) unit cell of germanene placed on top of a (6 $\times$ 6) unit cell of a MoS$_2$ monolayer with a lattice constant $a_{MoS_{2}}$ = 3.18 \AA, which yields a contracted germanene lattice constant $a_{Ge}=3.82$ \AA. Subsequently we optimize the atomic structure by taking Van der Waals interactions into account and find an average equilibrium buckling parameter of germanene $\Delta = 0.86 \pm 0.10 $ \AA \quad and a corresponding interlayer distance $d = 5.02$ \AA, defined as the distance between the in-plane averaged centers of mass of germanene and MoS$_2$.
 
The density of states (DOS) is depicted in Figure \ref{fig4} and shows a pronounced V-shape in the energy range of $\sim$1 eV near the Fermi energy stemming from electronic states of germanium. Moreover, the shape of both germanene and MoS$_2$ DOS are found to be in very good agreement with the experimental \textit{dI/dV} spectra shown in Figure \ref{fig2}(d). Moreover, the non-zero DOS at the Dirac point observed in experimental spectra is also reproduced. The only difference concerns the position of the Fermi energy, which is shifted in the calculated DOS toward higher energies by $\sim$0.3 eV. This shift indicates an $n$-type doping of the system. The absence of such a shift in the experimental spectra can be associated with the presence of acceptor impurities (e.g., O) or unsaturated defects in the sample.

\begin{figure}
\includegraphics[width=8.5cm]{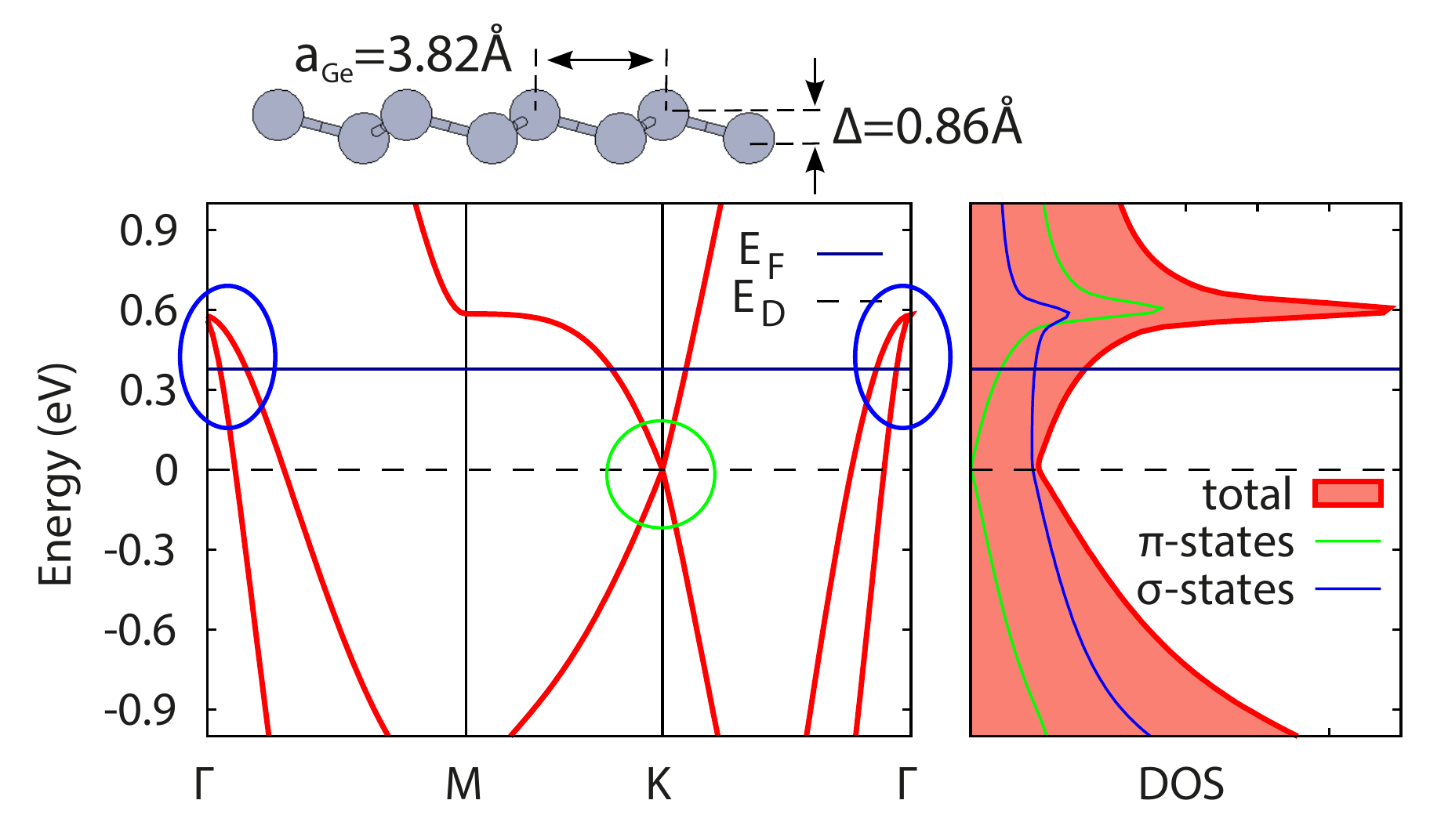}
\caption{
(Color on-line). Band structure shown along the high-symmetry points of the Brillouin zone and the corresponding orbital-resolved density of states calculated for contracted germanene with a lattice constant $a_{Ge}=3.82$ \AA and a buckling parameter $\Delta = 0.86 \pm 0.10 $ \AA. Horizontal solid and dashed lines correspond to the Fermi energy and Dirac point, respectively. Green and blue circles mark the $\pi$- and $\sigma$-bands, respectively.}
\label{fig5}
\end{figure}

To gain more insight into the origin of the finite DOS at the Dirac point as well as the observed doping, we calculate the band structure and perform orbital analysis of contracted free-standing germanene with the above mentioned structural parameters $a_{Ge}$ and $\Delta$. The results are shown in Figure \ref{fig5}. In addition to the $\pi$-bands leading to the formation of a Dirac cone in the vicinity of the K-point, there are two $\sigma$-bands close to the $\Gamma$-point. The emergence of those bands is related to the relatively large buckling of germanene, which makes the occupation of $\sigma$-states energetically less favorable \cite{twintig4} and leads to a $\sigma$-$\pi$ charge transfer. As a consequence of this process, the Fermi energy is shifted relative to the Dirac point, whereas unoccupied $\sigma$-bands produce a finite DOS in the relevant energy region. The $\sigma$-states contribution to the DOS is weakly dependent on energy as the corresponding dispersion is nearly quadratic ($E_{\sigma}(k) \sim k^{2}$). Therefore, we conclude that the observed V-shape of the total DOS originates predominantly from the $\pi$-states of germanene.

After having performed the STM and STS measurements we took the MoS$_2$/germanene sample out of the ultra-high vacuum system for an ex-situ analysis. X-ray photoelectron spectroscopy measurements revealed that the germanene has been oxidized. As a final remark we want to point out that hexagonal boron nitride is another very appealing substrate for the synthesis of germanene \cite{twintig7}. Hexagonal boron nitride has a band gap of 5.9 eV and a nearly perfect lattice match with germanene.

In summary, we have synthesized large continuous layers of germanene on a band gap material. The germanene islands preferentially nucleate at pre-existing defects of the MoS$_2$ surface. Germanene's lattice constant is about 20 \% larger than that of MoS$_2$ and the angle between the two lattices is 0 degrees. The density of states of the germanene layer exhibits a well-defined V-shape around the Fermi level, which hints to a 2D Dirac system. Unfortunately, the buckled germanene layer also has two parabolic bands that cross the Fermi level at the $\Gamma$ point. These states might suppress the anomalous quantum Hall effect as well as the 2D Dirac transport properties.

\begin{acknowledgments}
LZ and QY thank the China Scholarship Council for financial support. PB thanks the Nederlandse Organisatie voor Wetenschappelijk Onderzoek (NWO, STW 11431) for financial support. HJWZ and MIK thank the stichting voor Fundamenteel Onderzoek der Materie (FOM, FV157) for financial support. ANR and MIK acknowledge financial support by the European Union Seventh Framework Programme under Grant Agreement No. 604391 Graphene Flagship.
\end{acknowledgments}

\end{document}